\documentclass[12pt,a4paper]{JHEP3}
\usepackage{graphicx}
\usepackage{axodraw}
\usepackage{cite}
\newlength{\extra}
\title{Super-leading logarithms in non-global observables in QCD: Fixed order calculation}
\author{James Keates\\School of Physics \& Astronomy, University of Manchester,\\Oxford Road, Manchester, M13 9PL, U.K.\\\email{james.keates@cern.ch}}
\author{Michael H. Seymour\\School of Physics \& Astronomy, University of Manchester and\\ Theoretical Physics group (PH-TH), CERN, CH-1211 Geneva 23, Switzerland\\\email{mike.seymour@cern.ch}}
\preprint{MAN/HEP/2009/7\\CERN--PH--TH/2009--018}
\keywords{QCD,Jets}

\abstract{Super-leading logarithmic terms have previously been observed in non-global QCD observables. In this paper we re-calculate the first super-leading contribution to the `gaps between jets' cross-section using a diagrammatic fixed order approach. We calculate for the first time the super-leading contribution at fifth order in $\alpha_s$.}
\begin{document}


  \section{Introduction}

  Hard QCD interactions are typically accompanied by a large number of soft and collinear gluon emissions. These give rise to singularities, which cancel between real and virtual emissions, leaving a finite remainder. However when the real emission into a certain phase space region is restricted, this cancellation no longer occurs, and we are left with large logarithmic corrections to the hard process.
  
  For the processes we are interested in, the leading behaviour is obtained using the limit of soft, wide-angle (i.e.\ non-collinear) gluons. We can therefore  work in the Eikonal approximation. For virtual gluons in the Eikonal limit, we may integrate out one component of 4-momentum. This contour integral encloses a pole where the virtual gluon goes on-shell, and once integrated the remaining 3-momentum integral is of the exact form of a real gluon phase space integral, but with opposite sign. In an inclusive observable the cancellation between the real and virtual gluons in this limit is exact at the differential level. It therefore makes sense to formulate the calculation of this virtual gluon contribution as a phase space integral, and in particular to assign them rapidities corresponding to the rapidities of the real gluons they cancel against. We will therefore talk of virtual gluons occupying certain regions of phase space, meaning the limits on the remaining integrals over the virtual gluon momentum.
    
  An example of a restricted emission process is gaps-between-jets, where we explicitly limit the transverse energy emitted between hard di-jets. Traditionally \cite{Kidonakis:1998nf,Oderda:1998en,Oderda:1999kr,Berger:2001ns,Appleby:2003sj} it was assumed that the real--virtual cancellation was perfect outside the gap so that one need only consider the virtual gluons lying in the gap with $E_T$ above the transverse energy cut-off $Q_0$, the region in which real emission is vetoed and therefore cancellation is spoiled. These terms could be re-summed to all orders, to give a correction to the hard cross section. 
  
  However Dasgupta and Salam \cite{Dasgupta:2001sh,Dasgupta:2002bw} realised that one must also consider secondary emissions. Real gluons lying outside of the gap may emit back into the gap, and these emissions too are vetoed above $Q_0$. These give rise to a new tower of logarithms, formally as important as the previous primary emissions, known as non-global logarithms. One must now consider any number of emissions, real or virtual, outside the gap regions. The colour structure for this quickly becomes uncalculable and hence corrections have been calculated to all orders only in the large $N$ limit.
  
  Considering the specific case of only one gluon outside the gap, dressed with virtual gluons, but using the exact $N$ structure, Forshaw, Kyrieleis and Seymour \cite{Forshaw:2006fk} discovered a new class of logarithm. These logarithms appear first at order $\alpha_s^4$ relative to the hard process and are formally more important than the primary emission logarithm at this order, and therefore  known as super-leading logarithms. They are however sub-leading in colour, hence were not observed with the original Dasgupta-Salam non-global logarithms. Their origin is traced to a breakdown of colour coherence in the initial state originating from the imaginary part of loop integrals, known as Coulomb gluons. They are therefore only present in processes with at least two initial-state partons and at least two final-state partons. The calculation of \cite{Forshaw:2006fk} is limited by the requirement to calculate increasingly large anomalous dimension matrices as the number of external coloured partons increases, though recent steps have been made in simplifying it \cite{Forshaw:2008cq} using a colour basis independent approach \cite{Catani:1985xt,Bonciani:2003nt}.
  
  In this paper we present a diagrammatic approach to calculating the gaps-between-jets cross-section at fixed order. This allows us to retain the exact $N$ structure and consider any number of gluons outside the gap. The order that can be reached is limited by the large number of diagrams involved, and ultimately by the availability of computer power and time.
  
  In Section~\ref{FOchap} the diagrammatic algorithm is explained. Appendix A contains some details of the implementation of this algorithm in $C++$. In Section~\ref{reschap} we present results and in Section~\ref{outlookchap} an outlook.
  
  \section{Fixed Order Calculation}\label{FOchap}
  \subsection{Overview and Notation}\label{OverNote}
\FIGURE[hb]{
\centering
\SetWidth{2}
\SetScale{1}
\begin{centering}
\resizebox{4in}{!}{
\begin{picture}(520,210)(0,110)
\SetColor{Black}
%
\Gluon(60,260)(60,200){4}{4}
\Gluon(180,260)(180,200){4}{4}
\ArrowLine(0,280)(60,260)
\ArrowLine(60,260)(120,280)
\ArrowLine(120,280)(180,260)
\ArrowLine(180,260)(240,280)
\ArrowLine(0,180)(60,200)
\ArrowLine(60,200)(120,180)
\ArrowLine(120,180)(180,200)
\ArrowLine(180,200)(240,180)
\ArrowLine(120,320)(0,280)
\ArrowLine(240,280)(120,320)
\ArrowLine(120,140)(0,180)
\ArrowLine(240,180)(120,140)
\SetColor{Red}
\DashLine(120,130)(120,325){5}
\SetColor{Black}
\Gluon(330,260)(330,200){4}{4}
\Gluon(460,260)(460,200){4}{4}
\ArrowLine(270,280)(330,260)
\ArrowLine(330,260)(390,280)
\ArrowLine(400,280)(460,260)
\ArrowLine(460,260)(520,280)
\ArrowLine(270,180)(330,200)
\ArrowLine(330,200)(390,180)
\ArrowLine(400,180)(460,200)
\ArrowLine(460,200)(520,180)
\SetColor{Black}

\begin{LARGE}
\put(250,230){$=$}
\put(395,230){.}
\put(110,110){$\mathcal{|M|}^2$}
\put(320,110){$\mathcal{M}$}
\put(450,110){$\mathcal{M^*}$}
\end{LARGE}
\begin{Large}
\put(270,265){1}
\put(385,265){3}
\put(270,187){2}
\put(385,187){4}
\put(400,265){3}
\put(515,265){1}
\put(400,187){4}
\put(515,187){2}
\put(20,257){1}
\put(130,260){3}
\put(20,195){2}
\put(130,191){4}
\end{Large}

%
\end{picture}
}
\end{centering}
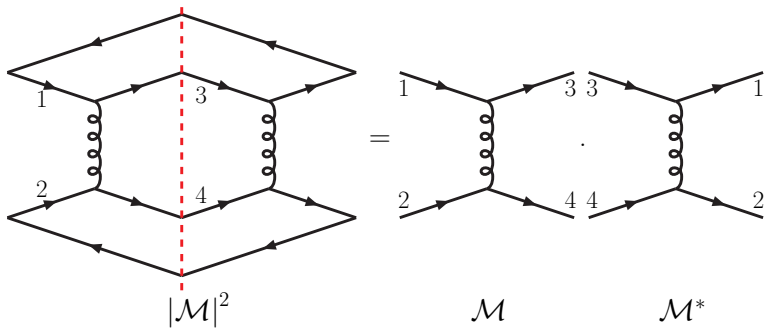
\caption{A cut diagram represents a squared matrix element.}
\label{msquare}
}
The calculation uses a representation of diagrams. These diagrams may be either `cut' or `uncut' with cut diagrams representing squared matrix elements, and uncut diagrams representing a set of cut diagrams. Figure~\ref{msquare} shows how a cut diagram corresponds to the matrix element squared formed from a matrix element and its complex conjugate. Figure~\ref{fourcuts} shows an uncut diagram, representing the sum of all possible cuts. A gluon running over a cut represents the interference of two real gluon emission diagrams. A gluon to one side of a cut represents a virtual gluon. All cuts through a diagram share the same colour factor (up to a possible overall sign, which we discuss in detail later).
\FIGURE[b]{
\centering
\SetWidth{2}
\SetScale{0.4}
\begin{centering}
\resizebox{\columnwidth}{!}{
\begin{picture}(520,74)(0,0)
\SetColor{Black}
%
\Gluon(60,125)(60,65){4}{4}
\Gluon(180,125)(180,65){4}{4}
\ArrowLine(0,145)(60,125)
\ArrowLine(60,125)(120,145)
\ArrowLine(120,145)(180,125)
\ArrowLine(180,125)(240,145)
\ArrowLine(0,45)(60,65)
\ArrowLine(60,65)(120,45)
\ArrowLine(120,45)(180,65)
\ArrowLine(180,65)(240,45)
\ArrowLine(120,185)(0,145)
\ArrowLine(240,145)(120,185)
\ArrowLine(120,5)(0,45)
\ArrowLine(240,45)(120,5)
\SetColor{Red}
\SetColor{Green}
\Gluon(120,145)(120,45){4}{6}
\SetColor{Black}
\Gluon(340,125)(340,65){4}{4}
\Gluon(460,125)(460,65){4}{4}
\ArrowLine(280,145)(340,125)
\ArrowLine(340,125)(400,145)
\ArrowLine(400,145)(460,125)
\ArrowLine(460,125)(520,145)
\ArrowLine(280,45)(340,65)
\ArrowLine(340,65)(400,45)
\ArrowLine(400,45)(460,65)
\ArrowLine(460,65)(520,45)
\ArrowLine(400,185)(280,145)
\ArrowLine(520,145)(400,185)
\ArrowLine(400,5)(280,45)
\ArrowLine(520,45)(400,5)
\SetColor{Red}
\DashLine(375,0)(375,190){5}
\SetColor{Green}
\Gluon(400,145)(400,45){4}{6}
\SetColor{Black}
\Gluon(600,125)(600,65){4}{4}
\Gluon(720,125)(720,65){4}{4}
\ArrowLine(540,145)(600,125)
\ArrowLine(600,125)(660,145)
\ArrowLine(660,145)(720,125)
\ArrowLine(720,125)(780,145)
\ArrowLine(540,45)(600,65)
\ArrowLine(600,65)(660,45)
\ArrowLine(660,45)(720,65)
\ArrowLine(720,65)(780,45)
\ArrowLine(660,185)(540,145)
\ArrowLine(780,145)(660,185)
\ArrowLine(660,5)(540,45)
\ArrowLine(780,45)(660,5)
\SetColor{Red}
\DashLine(635,0)(685,190){5}
\SetColor{Green}
\Gluon(660,145)(660,45){4}{6}
\SetColor{Black}
\Gluon(860,125)(860,65){4}{4}
\Gluon(980,125)(980,65){4}{4}
\ArrowLine(800,145)(860,125)
\ArrowLine(860,125)(920,145)
\ArrowLine(920,145)(980,125)
\ArrowLine(980,125)(1040,145)
\ArrowLine(800,45)(860,65)
\ArrowLine(860,65)(920,45)
\ArrowLine(920,45)(980,65)
\ArrowLine(980,65)(1040,45)
\ArrowLine(920,185)(800,145)
\ArrowLine(1040,145)(920,185)
\ArrowLine(920,5)(800,45)
\ArrowLine(1040,45)(920,5)
\SetColor{Red}
\DashLine(945,0)(895,190){5}
\SetColor{Green}
\Gluon(920,145)(920,45){4}{6}
\SetColor{Black}
\Gluon(1120,125)(1120,65){4}{4}
\Gluon(1240,125)(1240,65){4}{4}
\ArrowLine(1060,145)(1120,125)
\ArrowLine(1120,125)(1180,145)
\ArrowLine(1180,145)(1240,125)
\ArrowLine(1240,125)(1300,145)
\ArrowLine(1060,45)(1120,65)
\ArrowLine(1120,65)(1180,45)
\ArrowLine(1180,45)(1240,65)
\ArrowLine(1240,65)(1300,45)
\ArrowLine(1180,185)(1060,145)
\ArrowLine(1300,145)(1180,185)
\ArrowLine(1180,5)(1060,45)
\ArrowLine(1300,45)(1180,5)
\SetColor{Red}
\DashLine(1205,0)(1205,190){5}
\SetColor{Green}
\Gluon(1180,145)(1180,45){4}{6}
\SetColor{Black}

\begin{Large}
\put(99,36){$=$}
\put(208,36){$+$}
\put(312,36){$+$}
\put(417,36){$+$}
\end{Large}

%
\end{picture}
}
\end{centering}
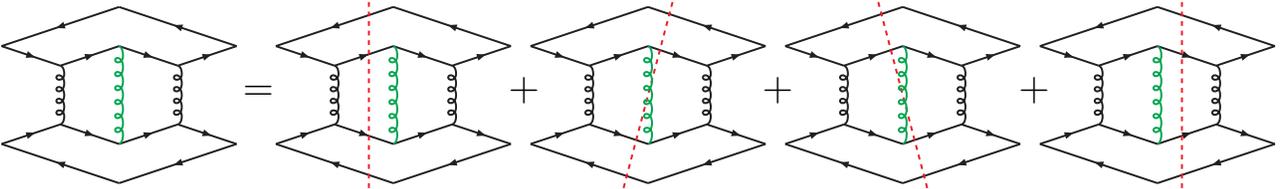
\caption{An uncut diagram represents a sum of cut diagrams.}
\label{fourcuts}
}

\newcommand{\slsh}{\rlap{$\;\!\!\not$}}

We work in the Eikonal (soft) gluon approximation, in which all components of the gluon momenta $k_i$ are much less than any components of the momenta $p_i$ of the original hard partons. In this limit, one need only consider Feynman diagrams in which the soft gluons are connected to the external hard partons, and not to internal lines within the hard scattering. Replacements of the form
\begin{equation}\label{eikonal}
\frac{\gamma^\mu (\slsh p+\slsh k)}{(p+k)^2 +i\epsilon}
\to \frac{p^\mu}{p.k+i\epsilon}
\end{equation}
are made in the expression for the matrix element. Equation~(\ref{eikonal}) represents the emission of a gluon of momentum $k$ from a quark of momentum $p$, however the form of the right-hand side is universal whether the parton is a quark or gluon. In the squared matrix element for gluon emission, where there are Eikonal attachments at both ends of the gluon (or in the matrix element for virtual gluons where this is always true), all gluons give terms of the form (in Feynman gauge)
\begin{equation}\label{eikonalsq}
\frac{p_i . p_j}{(p_i.k\pm i\epsilon)(p_j.k\pm i\epsilon)}
\end{equation}
where $p_i$ and $p_j$ are the momenta of two partons joined by a gluon of momentum $k$. The $\pm$ in the denominator depends on whether the gluon is emitted or absorbed, and whether the emitters sit in the initial or final state. For triple gluon vertices a further complication exists due to the anti-symmetric nature of the colour- and momentum-dependent parts of the triple gluon coupling. In order to correspond to our convention (the right hand side of eq.~(\ref{eikonal})) the soft gluon must be represented as being emitted to the right as one looks in the direction of momentum of a harder emitting gluon.

We retain terms leading in the logarithm of $Q$, the energy exchanged in the hard process, over the cut-off $Q_0$; the terms are also leading in $Y$, the rapidity separation, or $\pi^2$, which arises from Coulomb gluon terms. This is referred to in \cite{Forshaw:2005sx} as the $\pi^2$ Double Leading Logarithm Approximation ($\pi^2DLLA$). The colour basis independent approach of \cite{Forshaw:2008cq} shows that the real part of sub-leading terms in $Y$ in the anomalous dimension matrix are proportional to the identity matrix, and that this approximation contains all the interesting physics, in particular the super-leading logarithmic terms that interest us. 

The leading logarithms are obtained using the strong ordering condition. This states that each emission is much softer, in some chosen kinematic variable, e.g. $k_T$, than the last and can therefore be treated as independent. Each emission sees the result of previous emissions as the hard process and can therefore be calculated knowing only the momenta and colour configuration of the external partons. Previously added virtual gluons simply alter the colour structure of the hard process, real gluons mean the new emission is added, e.g.\ transforming a hard $2\to2$ process to a $2\to3$ process. The addition of each soft gluon to an external hard parton leads to a logarithmic integral over the internal hard parton propagator. The nesting of multiple emission results in a leading contribution with one logarithm for each order of $\alpha_s$.

Colour coherence would mean that the choice of ordering variable is unimportant: strong ordering in any kinematic variable of the soft gluons would give the same result. However as the appearance of super-leading logarithms is interpreted as a breakdown of colour coherence, the variable used may be crucial. In common with previous calculations, we use the transverse momentum as our ordering variable ($Q\gg k_{T,1}\gg k_{T,2}\gg ...\gg k_{T,n}\gg \Lambda_{QCD}$). This is, however, an area of on-going investigation.

It should be noted that, in the various approximations we use, all cuts through the softest gluon, as in figure~\ref{fourcuts}, cancel. When a phase space restriction is introduced this cancellation fails. This only applies for the softest gluon~-- when there is more than one gluon, complicated colour structures can ruin this cancellation, and one must consider all cuts through all diagrams.

By including all topologically distinct uncut diagrams, which include all allowed cuts, we can build up the basis of the entire order $\alpha_S$ correction to the hard process.

\subsection{Forming Uncut Diagrams}
The calculation starts with an original hard process (e.g.\ $qq\to qq$ exchanging a gluon). The hard process is formed into an uncut diagram, that is,  combined with its complex conjugate to form a closed loop. Only when this diagram is `cut', i.e. separated into a matrix element and complex conjugate pair can it be calculated physically. 

 From these uncut diagrams higher orders are built by iteratively adding gluons attaching to all possible external legs and previously added gluons. Real gluons may re-emit a gluon, although coupling to virtual gluons (including any in the initial hard process) is suppressed. In the Eikonal approximation using Feynman gauge, gluons attached at both ends to the same leg make no contribution, so these are not considered. 
 
 The colour factors corresponding to these diagrams are calculated at this stage. Here we must choose a representation of the triple gluon vertex, which may introduce a factor of $-1$ once the momentum direction of the partons is set later on. This process of adding gluons leaves us with a set of uncut diagrams each with a colour factor, a polynomial in the number of colours $N$. Figure~\ref{firstorder} shows all the possible first order uncut diagrams attaching to a $qq\to qq$ process.
\FIGURE[t]{
\centering
\SetWidth{2}
\SetScale{0.3}
\begin{centering}
\resizebox{\columnwidth}{!}{
\begin{picture}(468,54)(0,0)
\SetColor{Black}
%
\Gluon(80,120)(80,60){4}{4}
\Gluon(200,120)(200,60){4}{4}
\ArrowLine(20,140)(80,120)
\ArrowLine(80,120)(140,140)
\ArrowLine(140,140)(200,120)
\ArrowLine(200,120)(260,140)
\ArrowLine(20,40)(80,60)
\ArrowLine(80,60)(140,40)
\ArrowLine(140,40)(200,60)
\ArrowLine(200,60)(260,40)
\ArrowLine(140,180)(20,140)
\ArrowLine(260,140)(140,180)
\ArrowLine(140,0)(20,40)
\ArrowLine(260,40)(140,0)
\SetColor{Red}
\SetColor{Green}
\Gluon(140,140)(140,40){4}{6}
\SetColor{Black}
\Gluon(340,120)(340,60){4}{4}
\Gluon(460,120)(460,60){4}{4}
\ArrowLine(280,140)(340,120)
\ArrowLine(340,120)(400,140)
\ArrowLine(400,140)(460,120)
\ArrowLine(460,120)(520,140)
\ArrowLine(280,40)(340,60)
\ArrowLine(340,60)(400,40)
\ArrowLine(400,40)(460,60)
\ArrowLine(460,60)(520,40)
\ArrowLine(400,180)(280,140)
\ArrowLine(520,140)(400,180)
\ArrowLine(400,0)(280,40)
\ArrowLine(520,40)(400,0)
\SetColor{Red}
\SetColor{Green}
\Gluon(400,140)(400,180){4}{3}
\SetColor{Black}
\Gluon(600,120)(600,60){4}{4}
\Gluon(720,120)(720,60){4}{4}
\ArrowLine(540,140)(600,120)
\ArrowLine(600,120)(660,140)
\ArrowLine(660,140)(720,120)
\ArrowLine(720,120)(780,140)
\ArrowLine(540,40)(600,60)
\ArrowLine(600,60)(660,40)
\ArrowLine(660,40)(720,60)
\ArrowLine(720,60)(780,40)
\ArrowLine(660,180)(540,140)
\ArrowLine(780,140)(660,180)
\ArrowLine(660,0)(540,40)
\ArrowLine(780,40)(660,0)
\SetColor{Red}
\SetColor{Green}
\Gluon(660,0)(660,40){4}{3}
\SetColor{Black}
\Gluon(860,120)(860,60){4}{4}
\Gluon(980,120)(980,60){4}{4}
\ArrowLine(800,140)(860,120)
\ArrowLine(860,120)(920,140)
\ArrowLine(920,140)(980,120)
\ArrowLine(980,120)(1040,140)
\ArrowLine(800,40)(860,60)
\ArrowLine(860,60)(920,40)
\ArrowLine(920,40)(980,60)
\ArrowLine(980,60)(1040,40)
\ArrowLine(920,180)(800,140)
\ArrowLine(1040,140)(920,180)
\ArrowLine(920,0)(800,40)
\ArrowLine(1040,40)(920,0)
\SetColor{Red}
\SetColor{Green}
\Gluon(920,140)(920,0){4}{9}
\SetColor{Black}
\Gluon(1120,120)(1120,60){4}{4}
\Gluon(1240,120)(1240,60){4}{4}
\ArrowLine(1060,140)(1120,120)
\ArrowLine(1120,120)(1180,140)
\ArrowLine(1180,140)(1240,120)
\ArrowLine(1240,120)(1300,140)
\ArrowLine(1060,40)(1120,60)
\ArrowLine(1120,60)(1180,40)
\ArrowLine(1180,40)(1240,60)
\ArrowLine(1240,60)(1300,40)
\ArrowLine(1180,180)(1060,140)
\ArrowLine(1300,140)(1180,180)
\ArrowLine(1180,0)(1060,40)
\ArrowLine(1300,40)(1180,0)
\SetColor{Red}
\SetColor{Green}
\Gluon(1180,180)(1180,40){4}{9}
\SetColor{Black}
\Gluon(1380,120)(1380,60){4}{4}
\Gluon(1500,120)(1500,60){4}{4}
\ArrowLine(1320,140)(1380,120)
\ArrowLine(1380,120)(1440,140)
\ArrowLine(1440,140)(1500,120)
\ArrowLine(1500,120)(1560,140)
\ArrowLine(1320,40)(1380,60)
\ArrowLine(1380,60)(1440,40)
\ArrowLine(1440,40)(1500,60)
\ArrowLine(1500,60)(1560,40)
\ArrowLine(1440,180)(1320,140)
\ArrowLine(1560,140)(1440,180)
\ArrowLine(1440,0)(1320,40)
\ArrowLine(1560,40)(1440,0)
\SetColor{Red}
\SetColor{Green}
\Gluon(1440,180)(1440,0){4}{11}
\SetColor{Black}

%
\end{picture}
}
\end{centering}
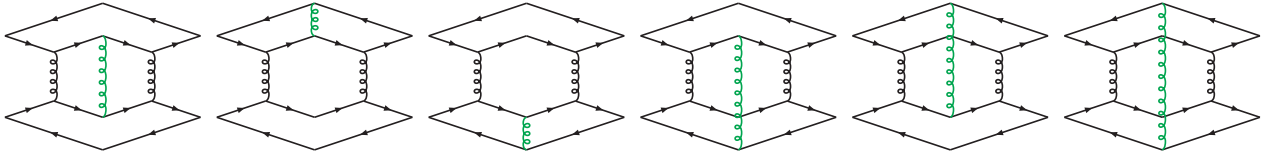
\caption{The set of all uncut first order diagrams.}
\label{firstorder}
}
  
  \subsection{Cutting and Ordering the Diagrams}
The next stage is to cut these diagrams. All possible cuts are considered. All original hard lines must be cut once  and only once, leaving a hard scatter either side of the cut. Vertices including added gluons, both those attached to hard partons and triple gluon vertices may sit either side of the cut.

 \FIGURE[t]{
\centering
\SetWidth{2}
\SetScale{0.4}
\begin{centering}
\resizebox{\columnwidth}{!}{
\begin{picture}(520,158)(0,0)
\SetColor{Black}
%
\Gluon(80,130)(80,70){4}{4}
\Gluon(200,130)(200,70){4}{4}
\ArrowLine(20,150)(80,130)
\ArrowLine(80,130)(140,150)
\ArrowLine(140,150)(200,130)
\ArrowLine(200,130)(260,150)
\ArrowLine(20,50)(80,70)
\ArrowLine(80,70)(140,50)
\ArrowLine(140,50)(200,70)
\ArrowLine(200,70)(260,50)
\ArrowLine(140,190)(20,150)
\ArrowLine(260,150)(140,190)
\ArrowLine(140,10)(20,50)
\ArrowLine(260,50)(140,10)
\SetColor{Green}
\Gluon(120,143)(120,57){4}{5}
\Gluon(160,143)(160,57){4}{5}
\SetColor{Red}
\DashLine(140,0)(140,195){5}
\SetColor{Black}
\Gluon(340,130)(340,70){4}{4}
\Gluon(460,130)(460,70){4}{4}
\ArrowLine(280,150)(340,130)
\ArrowLine(340,130)(400,150)
\ArrowLine(400,150)(460,130)
\ArrowLine(460,130)(520,150)
\ArrowLine(280,50)(340,70)
\ArrowLine(340,70)(400,50)
\ArrowLine(400,50)(460,70)
\ArrowLine(460,70)(520,50)
\ArrowLine(400,190)(280,150)
\ArrowLine(520,150)(400,190)
\ArrowLine(400,10)(280,50)
\ArrowLine(520,50)(400,10)
\SetColor{Green}
\Gluon(380,143)(380,57){4}{5}
\Gluon(420,143)(420,57){4}{5}
\SetColor{Red}
\DashLine(445,0)(395,195){5}
\SetColor{Black}
\Gluon(600,130)(600,70){4}{4}
\Gluon(720,130)(720,70){4}{4}
\ArrowLine(540,150)(600,130)
\ArrowLine(600,130)(660,150)
\ArrowLine(660,150)(720,130)
\ArrowLine(720,130)(780,150)
\ArrowLine(540,50)(600,70)
\ArrowLine(600,70)(660,50)
\ArrowLine(660,50)(720,70)
\ArrowLine(720,70)(780,50)
\ArrowLine(660,190)(540,150)
\ArrowLine(780,150)(660,190)
\ArrowLine(660,10)(540,50)
\ArrowLine(780,50)(660,10)
\SetColor{Green}
\Gluon(640,143)(640,57){4}{5}
\Gluon(680,143)(680,57){4}{5}
\SetColor{Red}
\DashLine(655,0)(705,195){5}
\SetColor{Black}
\Gluon(860,130)(860,70){4}{4}
\Gluon(980,130)(980,70){4}{4}
\ArrowLine(800,150)(860,130)
\ArrowLine(860,130)(920,150)
\ArrowLine(920,150)(980,130)
\ArrowLine(980,130)(1040,150)
\ArrowLine(800,50)(860,70)
\ArrowLine(860,70)(920,50)
\ArrowLine(920,50)(980,70)
\ArrowLine(980,70)(1040,50)
\ArrowLine(920,190)(800,150)
\ArrowLine(1040,150)(920,190)
\ArrowLine(920,10)(800,50)
\ArrowLine(1040,50)(920,10)
\SetColor{Green}
\Gluon(900,143)(900,57){4}{5}
\Gluon(940,143)(940,57){4}{5}
\SetColor{Red}
\DashLine(960,0)(960,195){5}
\SetColor{Black}
\Gluon(1120,130)(1120,70){4}{4}
\Gluon(1240,130)(1240,70){4}{4}
\ArrowLine(1060,150)(1120,130)
\ArrowLine(1120,130)(1180,150)
\ArrowLine(1180,150)(1240,130)
\ArrowLine(1240,130)(1300,150)
\ArrowLine(1060,50)(1120,70)
\ArrowLine(1120,70)(1180,50)
\ArrowLine(1180,50)(1240,70)
\ArrowLine(1240,70)(1300,50)
\ArrowLine(1180,190)(1060,150)
\ArrowLine(1300,150)(1180,190)
\ArrowLine(1180,10)(1060,50)
\ArrowLine(1300,50)(1180,10)
\SetColor{Green}
\Gluon(1160,143)(1160,57){4}{5}
\Gluon(1200,143)(1200,57){4}{5}
\SetColor{Red}
\DashLine(1100,0)(1260,195){5}
%
\SetColor{Black}
\Gluon(80,330)(80,270){4}{4}
\Gluon(200,330)(200,270){4}{4}
\ArrowLine(20,350)(80,330)
\ArrowLine(80,330)(140,350)
\ArrowLine(140,350)(200,330)
\ArrowLine(200,330)(260,350)
\ArrowLine(20,250)(80,270)
\ArrowLine(80,270)(140,250)
\ArrowLine(140,250)(200,270)
\ArrowLine(200,270)(260,250)
\ArrowLine(140,390)(20,350)
\ArrowLine(260,350)(140,390)
\ArrowLine(140,210)(20,250)
\ArrowLine(260,250)(140,210)
\SetColor{Green}
\Gluon(120,343)(120,257){4}{5}
\Gluon(160,343)(160,257){4}{5}
\SetColor{Red}
\DashLine(100,205)(100,395){5}
\SetColor{Black}
\Gluon(340,330)(340,270){4}{4}
\Gluon(460,330)(460,270){4}{4}
\ArrowLine(280,350)(340,330)
\ArrowLine(340,330)(400,350)
\ArrowLine(400,350)(460,330)
\ArrowLine(460,330)(520,350)
\ArrowLine(280,250)(340,270)
\ArrowLine(340,270)(400,250)
\ArrowLine(400,250)(460,270)
\ArrowLine(460,270)(520,250)
\ArrowLine(400,390)(280,350)
\ArrowLine(520,350)(400,390)
\ArrowLine(400,210)(280,250)
\ArrowLine(520,250)(400,210)
\SetColor{Green}
\Gluon(380,343)(380,257){4}{5}
\Gluon(420,343)(420,257){4}{5}
\SetColor{Red}
\DashLine(355,205)(405,395){5}
\SetColor{Black}
\Gluon(600,330)(600,270){4}{4}
\Gluon(720,330)(720,270){4}{4}
\ArrowLine(540,350)(600,330)
\ArrowLine(600,330)(660,350)
\ArrowLine(660,350)(720,330)
\ArrowLine(720,330)(780,350)
\ArrowLine(540,250)(600,270)
\ArrowLine(600,270)(660,250)
\ArrowLine(660,250)(720,270)
\ArrowLine(720,270)(780,250)
\ArrowLine(660,390)(540,350)
\ArrowLine(780,350)(660,390)
\ArrowLine(660,210)(540,250)
\ArrowLine(780,250)(660,210)
\SetColor{Green}
\Gluon(640,343)(640,257){4}{5}
\Gluon(680,343)(680,257){4}{5}
\SetColor{Red}
\DashLine(665,205)(615,395){5}
\SetColor{Black}
\Gluon(860,330)(860,270){4}{4}
\Gluon(980,330)(980,270){4}{4}
\ArrowLine(800,350)(860,330)
\ArrowLine(860,330)(920,350)
\ArrowLine(920,350)(980,330)
\ArrowLine(980,330)(1040,350)
\ArrowLine(800,250)(860,270)
\ArrowLine(860,270)(920,250)
\ArrowLine(920,250)(980,270)
\ArrowLine(980,270)(1040,250)
\ArrowLine(920,390)(800,350)
\ArrowLine(1040,350)(920,390)
\ArrowLine(920,210)(800,250)
\ArrowLine(1040,250)(920,210)
\SetColor{Green}
\Gluon(900,343)(900,257){4}{5}
\Gluon(940,343)(940,257){4}{5}
\SetColor{Red}
\DashLine(920,205)(920,395){5}
\SetColor{Black}
\Gluon(1120,330)(1120,270){4}{4}
\Gluon(1240,330)(1240,270){4}{4}
\ArrowLine(1060,350)(1120,330)
\ArrowLine(1120,330)(1180,350)
\ArrowLine(1180,350)(1240,330)
\ArrowLine(1240,330)(1300,350)
\ArrowLine(1060,250)(1120,270)
\ArrowLine(1120,270)(1180,240)
\ArrowLine(1180,250)(1240,270)
\ArrowLine(1240,270)(1300,250)
\ArrowLine(1180,390)(1060,350)
\ArrowLine(1300,350)(1180,390)
\ArrowLine(1180,210)(1060,250)
\ArrowLine(1300,250)(1180,210)
\SetColor{Green}
\Gluon(1160,343)(1160,257){4}{5}
\Gluon(1200,343)(1200,257){4}{5}
\SetColor{Red}
\DashLine(1260,205)(1100,395){5}
\SetColor{Black}
\begin{large}
\put(8,118){$a.)$}
\put(112,118){$b.)$}
\put(216,118){$c.)$}
\put(320,118){$d.)$}
\put(424,118){$e.)$}
\put(8,38){$f.)$}
\put(112,38){$g.)$}
\put(216,38){$h.)$}
\put(320,38){$i.)$}
\put(424,38){$j.)$}
\end{large}
%
\end{picture}
}
\end{centering}
\caption{All possible cuts through a second order diagram.}
\label{secondcut}
}
  
  The cut diagrams are subsequently ordered. We assume strong ordering, in order to capture the leading logarithms, as explained in section~\ref{OverNote}. All restrictions are obtained by moving outwards along the hard parton lines from the hard process to the cut, and similarly along added gluon lines. Diagrams with more than one possible ordering have all orderings retained separately, those which are inconsistent with strong ordering are rejected. Also rejected are cuts leading to disconnected gluons and those where soft gluons attach only to one harder parton (it can be seen from eq.~(\ref{eikonalsq}) that these terms will be 0 for massless partons). 
  
  \FIGURE[b]{
\centering
\SetWidth{2}
\SetScale{0.65}
\begin{centering}
\resizebox{5in}{!}{
\begin{picture}(540,123)(0,90)
\SetColor{Black}
%
\Gluon(60,260)(60,200){4}{4}
\Gluon(180,260)(180,200){4}{4}
\ArrowLine(0,280)(60,260)
\ArrowLine(60,260)(120,280)
\ArrowLine(120,280)(180,260)
\ArrowLine(180,260)(240,280)
\ArrowLine(0,180)(60,200)
\ArrowLine(60,200)(120,180)
\ArrowLine(120,180)(180,200)
\ArrowLine(180,200)(240,180)
\ArrowLine(120,320)(0,280)
\ArrowLine(240,280)(120,320)
\ArrowLine(120,140)(0,180)
\ArrowLine(240,180)(120,140)
\SetColor{Green}
\Gluon(138,274)(123,221){4}{4}
\Gluon(123,221)(75,155){4}{6}
\Gluon(123,221)(141,187){4}{3}
\SetColor{Black}
\Gluon(340,260)(340,200){4}{4}
\Gluon(460,260)(460,200){4}{4}
\ArrowLine(280,280)(340,260)
\ArrowLine(340,260)(400,280)
\ArrowLine(400,280)(460,260)
\ArrowLine(460,260)(520,280)
\ArrowLine(280,180)(340,200)
\ArrowLine(340,200)(400,180)
\ArrowLine(400,180)(460,200)
\ArrowLine(460,200)(520,180)
\ArrowLine(400,320)(280,280)
\ArrowLine(520,280)(400,320)
\ArrowLine(400,140)(280,180)
\ArrowLine(520,180)(400,140)
\SetColor{Red}
\DashLine(480,130)(360,325){5}
\SetColor{Blue}
\Gluon(418,274)(355,155){4}{10}
\SetColor{Green}
\Gluon(395,227)(421,187){4}{3}
\SetColor{Black}
\Gluon(600,260)(600,200){4}{4}
\Gluon(720,260)(720,200){4}{4}
\ArrowLine(540,280)(600,260)
\ArrowLine(600,260)(660,280)
\ArrowLine(660,280)(720,260)
\ArrowLine(720,260)(780,280)
\ArrowLine(540,180)(600,200)
\ArrowLine(600,200)(660,180)
\ArrowLine(660,180)(720,200)
\ArrowLine(720,200)(780,180)
\ArrowLine(660,320)(540,280)
\ArrowLine(780,280)(660,320)
\ArrowLine(660,140)(540,180)
\ArrowLine(780,180)(660,140)
\SetColor{Red}
\DashLine(750,130)(630,325){5}
\SetColor{Blue}
\Gluon(678,274)(681,187){-4}{7}
\SetColor{Green}
\Gluon(615,155)(680,227){4}{8}
\SetColor{Black}
\begin{Large}
\put(165,147){=}
\put(341,147){+}
\put(509,147){+ ...}
\end{Large}
\begin{large}
\put(63,134){1}
\put(73,155){2}
\put(90,134){3}
\put(238,134){1}
\put(250,157){2}
\put(272,134){3}
\put(418,134){1}
\put(429,155){2}
\put(448,134){3}
\end{large}

%
\end{picture}
}
\end{centering}
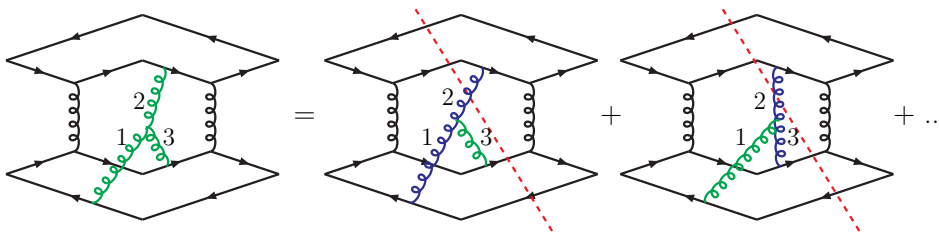
\caption{A triple gluon diagram and two possible orderings arising from one cut.}
\label{TGVpic}
}

  Figure~\ref{secondcut} shows all possible cuts through a second order diagram. Note that diagrams d.) and f.) are the same, but two possible orderings exist so we keep a diagram for each. We can see in diagrams a.) to c.) the leftmost gluon must be softest, as it is furthest from the hard process to the right of the cut. We might also notice that diagrams a.) to d.) resemble figure~\ref{fourcuts}, and we are free to assume the leftmost gluon can be softest in diagram d.) as well. This set forms a cancelling set of all cuts through the softest gluon. In order to include all possible orderings diagram f.) must be the alternative ordering of d.) with the rightmost gluon being softest, and forming a canceling set with g.) to i.).  To the left of the cut in e.) the leftmost gluon is closest to the hard process and therefore hardest. But to the right, the rightmost gluon is hardest. This arrangement is incompatible with the strong ordering we assume. Diagrams e.) and j.) must therefore be rejected.
  
   At this stage triple gluon vertices are also considered~-- the emitting gluon must be real, and harder than the emitted gluon and we impose these conditions during the ordering phase. Three gluons meeting at a vertex, treated equally when obtaining unique uncut diagrams now become one hard gluon emitting one softer gluon. The ordering stage determines the momenta of the gluons, and we may now resolve the factor of $-1$ ambiguity resulting from making a specific choice of order in the triple gluon coupling, $if_{ijk}$ when we calculated the colour factors of the uncut diagrams\label{minus1b}.
  
  For example in figure~\ref{TGVpic}, the colour factor is calculated at the uncut diagram stage, shown on the left, representing the triple gluon vertex as $i f_{123}$. The first cut diagram shows one possible ordering with gluon 1-2 emitting gluon 3. Momentum flows from the emission of gluon 1, and gluon 3 is emitted to the right as one looks along the momentum direction, the convention we have adopted for the Eikonal approximation. The second cut diagram shows an alternative ordering of the same cut, with gluon 3-2 being hard and gluon 1 soft. Momentum flows from the emission of gluon 3 upward, and gluon 2 is emitted to the left of the momentum direction. As our convention requires the colour factor to be written as if the gluon is emitted to the right, in this case, $i f_{213}$, we acquire an overall minus sign for this second cut diagram, which is taken into the colour factor.
  
  \subsection{Forming the Integrals}\label{intsect}
  
  Now we have all the information needed to perform the calculation, a set of integrals over the added gluon four momenta is formed from these ordered diagrams. Each integral is defined only by the gluon whose momentum we integrate over, the two partons the gluon stretches between (of the form in eq.~(\ref{eikonalsq})) and whether the gluon is real or virtual. For example, if we take the cut diagram b.) from figure~\ref{secondcut} we have a harder virtual gluon, $k_1$, and a softer real gluon, $k_2$, and from this we form:
 \begin{eqnarray}\label{sigma1}
\hspace{-1cm}\sigma=\frac{\sigma_0}{C_0 (N)} C (N) g_s^4\int_{virt}\frac{p_3.p_4}{(p_3.k_1+i\epsilon)(p_4.k_1-i\epsilon)}\frac{i}{k_1^2+i\epsilon}\frac{\mathrm{d}^4k_1}{(2\pi)^4}
\int_{real}\frac{p_3.p_4}{(p_3.k_2)(p_4.k_2)}\frac{\mathrm{d}^3k_2}{2k_2^0(2\pi)^3}\hspace{-1cm}
\nonumber\\
\end{eqnarray}
where $\sigma_0$ and $C_0$ are the cross section and colour factor of the original hard process, $C$ the colour factor of the new process and  $p_3$ and $p_4$ the momenta of the two outgoing quarks. 

Writing the momenta in the centre of mass frame of the incoming partons, we get
\begin{eqnarray}
p_1 & = & \frac{\sqrt{s}}{2}(1;0,0,1) \\
p_2 & = & \frac{\sqrt{s}}{2}(1;0,0,-1) \\
p_3 & = & Q\left(\cosh(\delta \eta/2);0,1,\sinh(\delta\eta/2)\right) \\
p_4 & = & Q\left(\cosh(\delta \eta/2);0,-1,-\sinh(\delta\eta/2)\right) \\
k_1 & = & k_{T,1}\left(\cosh(y_1);\sin(\phi_1),\cos(\phi_1),\sinh(y_1)\right) \\
k_2 & = & k_{T,2}\left(\cosh(y_2);\sin(\phi_2),\cos(\phi_2),\sinh(y_2)\right) 
\end{eqnarray}
 The $dk^0_1$ integral in the virtual gluon can be evaluated as a complex integral and gives two poles, one corresponding to gluon 1 being on-shell (known as the eikonal gluon), and one corresponding to an emitting parton being on-shell (known as the Coulomb gluon) giving an imaginary term $i\pi$ if the emitters are both in the initial or both in the final state (as they are in this example). After performing the azimuthal ($\phi$) integrals it can be shown that the rapidity dependent integrands are of the form $\frac{e^y}{2\sinh(|y|)}$ and in the large $|y|$ limit reduce\footnote{There is a subtlety here: the original expression is divergent at the point $y=0$, corresponding to emission collinear with one of the hard partons, while the approximated expression is finite (although discontinuous) there. The approximation is justified by the fact, proved using the colour basis independent notation in \cite{Forshaw:2008cq}, that final-state collinear divergences cancel to all orders and so do not have an effect on the final result. The large $Y$ limit is therefore captured by this step function approximation.} to a step function $\theta(y)$. This leaves eq.~(\ref{sigma1}) as
 \begin{eqnarray}\label{sigma2}
\hspace{-1cm}\sigma=\frac{\sigma_0}{C_0 (N)} C (N) \left(\frac{\alpha_s}{\pi}\right)^2\int_{Q_0}^Q \frac{dk_{T,1}}{k_{T,1}} \int_{Q_0}^{k_1} \frac{dk_{T,2}}{k_{T,2}} \left(\int_{-\frac{\delta\eta}{2}}^{\frac{\delta\eta}{2}} dy_1 -i\pi\right)\left(\int_{\frac{Y}{2}}^{\frac{\delta\eta}{2}}dy_2+\int^{-\frac{Y}{2}}_{-\frac{\delta\eta}{2}}dy_2\right)\hspace{-1cm}
\nonumber\\
\end{eqnarray}
where $Q_0$ is the energy cut-off into the gap, $k_{T,2}$ is cut off by $k_{T,1}$ as we have assumed strong $k_T$ ordering and the real gluon rapidity is restricted to the region outside the gap (see figure~\ref{etaphi}). In general these contributions come with an overall sign that must be determined from the colour factor, the number of cut gluons, the orientation of triple-gluon vertices, etc.

  As we can see, the rapidity integrals only depend on the limits. We also keep Coulomb gluon terms, which are the imaginary part of the virtual gluon integrals of the form $i\pi$. The physical observable is real, but may contain terms proportional to $\pi ^n$ arising from these terms, where $n$ is even.
    \FIGURE[ht]{
\label{etaphi}
\centering
\SetWidth{2}
\SetScale{1}
\begin{centering}
\resizebox{4in}{!}{
\begin{picture}(500,250)(0,-75)
\SetColor{Black}
%
\Boxc(250,50)(500,250)
\DashLine(150,175)(150,-75){5}
\DashLine(350,175)(350,-75){5}
\CCirc(150,100){25}{Black}{Red}
\CCirc(350,0){25}{Black}{Red}
\GBoxc(250,50)(150,250){0.8}
\LongArrow(240,150)(175,150)
\LongArrow(260,150)(325,150)
\Text(250,150)[]{$Y$}
\LongArrow(240,-50)(150,-50)
\LongArrow(260,-50)(350,-50)
\Text(250,-50)[]{$\delta\eta$}
\LongArrow(415,50)(350,50)
\LongArrow(435,50)(500,50)
\Text(425,50)[]{$\frac{\bar{Y}}{2}$}
\LongArrow(20,-55)(80,-55)
\LongArrow(20,-55)(20,5)
\Text(50,-63)[]{$\eta$}
\Text(12,-25)[]{$\phi$}

%
\end{picture}
}
\end{centering}
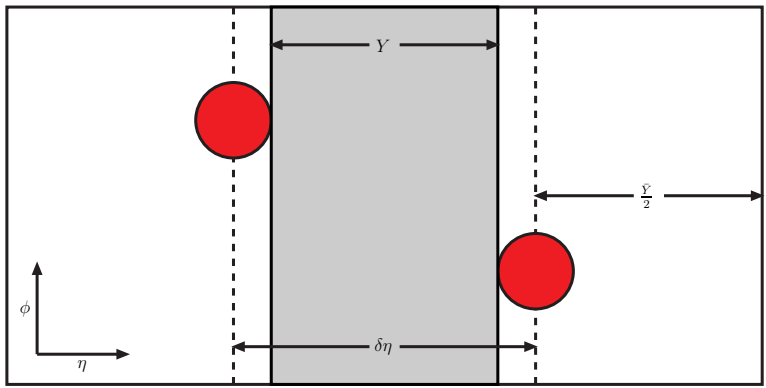
\caption{An $\eta$-$\phi$ plot showing the jets (red circles) and gap (grey band).}
}

We notice only the colour factor and rapidity integrals change between diagrams, the transverse momentum and numerical factors change only between orders in $\alpha_s$. After azimuthal averaging in the high energy (large $Y$) limit the rapidity of the gluon is restricted to lie between that of the two partons it connects.  Therefore, for each gluon, we get a rapidity dependent contribution dependent only on which region on the plot~\ref{etaphi} it occupies. Gluons coupled to other gluons form a set of nested integrals. Answers are given in terms of these rapidity zones, the width of the gap region, $Y$, the out-of-gap region, $\bar{Y}$, and the jet radius, $R=\frac{1}{2}(\delta\eta-Y)$.

Out of gap contributions ($\bar{Y}$ terms) cancel over the set of all diagrams up to order $\alpha_s^3$. When this cancellation fails, the out of gap region (with the rapidity of the incoming partons infinite) must be dealt with specially, as described in the next section.

\subsubsection{Super-Leading Logarithms}

The out-of-gap rapidity region appears to be infinite, but at large rapidities a collinear rather than soft approximation is more appropriate. Matching the soft and collinear regions in \cite{Forshaw:2006fk} it was argued that the out-of-gap rapidity interval should be cut off at $\bar{Y}=2\ln\left(\frac{Q}{k_T}\right)$. These are the super-leading logarithms, which must slot into the nested $k_T$ integrals (as seen in eq.~(\ref{sigma2})). 

We require an extra factor, depending on where in the $k_T$ ordered chain the gluon lies. With no out-of-gap contribution we get something of the form $\ln^n\left(\frac{Q}{Q_0}\right)/n!$. For each additional logarithm from an out-of-gap gluon, $n$ increases by one in both the power of the logarithm and the factorial, and there is an extra numerical factor of the number of harder logarithms plus one. For one gluon outside the gap at fourth order, for example, we have
  \begin{equation}\label{sleq}
\frac{\ln^4\left(\frac{Q}{Q_0}\right)}{4!}\bar{Y}\to\frac{2 \ln^5\left(\frac{Q}{Q_0}\right)}{5!}(m+1)
  \end{equation}
  where $m$ is the number of gluons harder than the out-of-gap (``super-leading") gluon. Similar equations hold for larger numbers of out-of-gap gluons.
  
  We can now re-create the full answer by taking the summed colour and rapidity dependent parts calculated on a diagram-by-diagram basis, and combining them with the $k_T$ logarithms (accounting for superleading logarithms), $\alpha_S$ and hard cross-section pieces calculated on an order-by-order basis. 
  
  We have written a $C++$ program to encode the calculation, in order to systematically ensure all diagrams and cuts were accounted for. Further details are contained in the appendix.
  
  \section{Results}\label{reschap}
  Here we present the results as corrections to $\sigma_0$, the hard scattering process with no additional soft gluons. The main original results are the two gluon outside the gap coefficients (eqs.~(\ref{s2qq}), (\ref{s2qg}), (\ref{s2gg}), (\ref{s2sqq})), but we show the full $\pi^2DLLA$ results at fourth and fifth order in $\alpha_s$, calculated at the same time for comparison. The one gluon outside the gap coefficients at fifth order, and for colour-singlet hard processes, are also new.
  
  We present here results for 0,1 and 2 gluons outside the gap. The calculation includes any number of gluons outside the gap, contributions from higher numbers of gluons outside cancel in the sum over diagrams. This explicitly shows, for example, that there is no two gluon outside contribution at $O(\alpha_s^4)$, as argued in \cite{Forshaw:2006fk}. 
  
  The results given are for the $t$-channel scattering of various particles exchanging a gluon, and for two quarks exchanging a colour singlet. Although the colour singlet case is subleading compared to the gluon exchange for simple scattering, it shares a colour structure with Higgs production through vector boson fusion, for which a rapidity gap is an important experimental signal, and which has been the subject of previous studies into soft gluon resummation \cite{Forshaw:2007vb}. Exchanging an anti-quark for a quark leaves the results unchanged.
  
  For quark-quark scattering, exchanging a gluon at order $\alpha_s^4$:
  \begin{eqnarray}\label{as4qq}
  \sigma_{0,qq,\alpha_s^4}&=&\sigma_0\left( \frac{2\alpha_s}{\pi} \right) ^4\ln^4\left(\frac{Q}{Q_0}\right)\left[ Y^4 \frac{N^4}{24}-\pi^2Y^2\frac{5(N^2-1)}{96}\right],\\
  \sigma_{1,qq,\alpha_s^4}&=&-\sigma_0\left( \frac{2\alpha_s}{\pi} \right) ^4\ln^5\left(\frac{Q}{Q_0}\right)\pi^2 Y\frac{3N^2-4}{240},
  \end{eqnarray}
  and at order $\alpha_s^5$:
  \begin{eqnarray}\label{as5qq}
  \nonumber\sigma_{0,qq,\alpha_s^5}&=&-\sigma_0\left( \frac{2\alpha_s}{\pi} \right) ^5\ln^5\left(\frac{Q}{Q_0}\right)\\
  &&\left [Y^5\frac{N^5}{120}-\pi^2Y^3\frac{17(N^3-N)}{960}+\pi^4Y\frac{(N^2-1)}{240N}\right],\\
  \sigma_{1,qq,\alpha_s^5}&=&\sigma_0\left( \frac{2\alpha_s}{\pi} \right) ^5\ln^6\left(\frac{Q}{Q_0}\right)\pi^2 Y^2\frac{8N^3-11N}{960},\\
  \sigma_{2,qq,\alpha_s^5}&=&\sigma_0\left( \frac{2\alpha_s}{\pi} \right) ^5\ln^7\left(\frac{Q}{Q_0}\right)\pi^2 Y\frac{27N^3-44N}{20160}.\label{s2qq}
  \end{eqnarray}
  For quark-gluon scattering we find:
  \begin{eqnarray}\label{as4qg}
  \sigma_{0,qg,\alpha_s^4}&=&\sigma_0\left( \frac{2\alpha_s}{\pi} \right) ^4\ln^4\left(\frac{Q}{Q_0}\right) \left[Y^4 \frac{N^4}{24}-\pi^2Y^2\frac{5N^2}{48}\right],\\
  \sigma_{1,qg,\alpha_s^4}&=&-\sigma_0\left( \frac{2\alpha_s}{\pi} \right) ^4\ln^5\left(\frac{Q}{Q_0}\right)\pi^2 Y\frac{7N^2}{240},
  \end{eqnarray}
  \begin{eqnarray}\label{as5qg}
  \nonumber\sigma_{0,qg,\alpha_s^5}&=&-\sigma_0\left( \frac{2\alpha_s}{\pi} \right) ^5\ln^5\left(\frac{Q}{Q_0}\right)\\
  &&\left[Y^5\frac{N^5}{120}-\pi^2Y^3\frac{17N^3}{480}+\pi^4Y\frac{5N^3+16N}{1920}\right],\\
  \sigma_{1,qg,\alpha_s^5}&=&\sigma_0\left( \frac{2\alpha_s}{\pi} \right) ^5\ln^6\left(\frac{Q}{Q_0}\right)\pi^2 Y^2\frac{10N^3+3N}{480},\\
  \sigma_{2,qg,\alpha_s^5}&=&\sigma_0\left( \frac{2\alpha_s}{\pi} \right) ^5\ln^7\left(\frac{Q}{Q_0}\right)\pi^2 Y\frac{37N^3+9N}{10080}.\label{s2qg}
  \end{eqnarray}
   For gluon-gluon scattering we obtain:
  \begin{eqnarray}\label{as4gg}
  \sigma_{0,gg,\alpha_s^4}&=&\sigma_0\left( \frac{2\alpha_s}{\pi} \right) ^4\ln^4\left(\frac{Q}{Q_0}\right) \left[Y^4 \frac{N^4}{24}-\pi^2Y^2\frac{N^2+36}{48}\right],\\
  \sigma_{1,gg,\alpha_s^4}&=&-\sigma_0\left( \frac{2\alpha_s}{\pi} \right) ^4\ln^5\left(\frac{Q}{Q_0}\right)\pi^2 Y\frac{3N^2+4}{80},
  \end{eqnarray}
  \begin{eqnarray}\label{as5gg}
 \nonumber \sigma_{0,gg,\alpha_s^5}&=&-\sigma_0\left( \frac{2\alpha_s}{\pi} \right) ^5\ln^5\left(\frac{Q}{Q_0}\right)\\
  &&\left[Y^5\frac{N^5}{120}+\pi^2Y^3\frac{43N^3+500N}{480}+\pi^4Y\frac{N^3+8N}{240}\right],\\
  \sigma_{1,gg,\alpha_s^5}&=&\sigma_0\left( \frac{2\alpha_s}{\pi} \right) ^5\ln^6\left(\frac{Q}{Q_0}\right)\pi^2 Y^2\frac{11N^3-28N}{960},\\
  \sigma_{2,gg,\alpha_s^5}&=&\sigma_0\left( \frac{2\alpha_s}{\pi} \right) ^5\ln^7\left(\frac{Q}{Q_0}\right)\pi^2 Y\frac{35N^3+108N}{6720}.\label{s2gg}
  \end{eqnarray}
  For quark-quark scattering exchanging a singlet:
  \begin{eqnarray}\label{as4sqq}
  \sigma_{0,qsq,\alpha_s^4}&=&\sigma_0\left( \frac{2\alpha_s}{\pi} \right) ^4\ln^4\left(\frac{Q}{Q_0}\right) \pi^2Y^2 \frac{N^2-1}{32},\\
  \sigma_{1,qsq,\alpha_s^4}&=&\sigma_0\left( \frac{2\alpha_s}{\pi} \right) ^4\ln^5\left(\frac{Q}{Q_0}\right)\pi^2 Y\frac{N^2-1}{120},
  \end{eqnarray}
   \begin{eqnarray}\label{as5sqq}
 \nonumber \sigma_{0,qsq,\alpha_s^5}&=&-\sigma_0\left( \frac{2\alpha_s}{\pi} \right) ^5\ln^5\left(\frac{Q}{Q_0}\right)\\
  &&\left[\pi^2Y^3\frac{7(N^3-N)}{960}-\pi^4Y\frac{N^2-1}{240N}\right],\\
  \sigma_{1,qsq,\alpha_s^5}&=&-\sigma_0\left( \frac{2\alpha_s}{\pi} \right) ^5\ln^6\left(\frac{Q}{Q_0}\right)\pi^2 Y^2\frac{N^3-N}{320},\\
  \sigma_{2,qsq,\alpha_s^5}&=&-\sigma_0\left( \frac{2\alpha_s}{\pi} \right) ^5\ln^7\left(\frac{Q}{Q_0}\right)\pi^2 Y\frac{N^3-N}{1260}.\label{s2sqq}
  \end{eqnarray}
  It is interesting to note that in each case the fifth-order one-outside and two-outside cross sections have the same sign, opposite to that of the fourth-order one-outside cross section.

  \subsection{Checks}
  A series of checks against pre-existing results were undertaken in order to validate the performance of the algorithm, and gain confidence in the original results.
  
  The number of unique uncut diagrams produced is in theory calculable, however it is very complex for the large numbers of diagrams we dealt with. However by isolating subsets of diagrams (e.g. for two or three hard parton states, for five gluon configurations containing seven triple gluon vertices (and therefore only three hard attachments) over four hard partons) and comparing with the \textsf{qgraf} program\cite{Nogueira:1991ex} we could be confident we had the full set of unique diagrams at each order.
  
  A major complication encountered in the completion of this work was the correct assignment of the $\pm1$ factors arising in triple gluon vertices. The failure of this aspect of the calculation gave rise to imaginary ($i\pi$) terms remaining in the final answers and we consider the eventual elimination of these terms a good test of our sign convention.
 
  Terms with two or more gluons outside the gap, although not explicitly calculated previously, have been argued to cancel below $O(\alpha_s^5)$ \cite{Forshaw:2006fk} and our results show this explicitly. We also find those with three or more gluons cancel at $O(\alpha_s^5)$ as expected.
  
  The leading $Y$ and $\pi$ terms reproduce the result of a ladder of virtual gluons lying in the gap and have been checked against other calculations at all orders. The one gluon outside the gap terms reproduce the results presented at order $\alpha_s^4$ in \cite{Forshaw:2008cq}, and have been compared for the quark-only processes at order $\alpha_s^5$ with a further expansion of the exponential approach of \cite{Forshaw:2006fk}.

  \section{Outlook}\label{outlookchap}
  In this paper we have presented a diagrammatic method for calculating large logarithmic corrections arising from the addition of soft gluons to a fixed order in $\alpha_s$. We have presented the first two gluon out of the gap corrections to the gaps-between-jets cross-section.  The results support the speculation of \cite{Forshaw:2006fk}, that at each higher order one obtains an additional logarithm from contributions with an additional gluon outside the gap, and hence that the true leading behaviour is not $\alpha_s^n\log^nQ/Q_0$, as once thought, but $\alpha_s^n\log^{2n-3}Q/Q_0$.  Clearly to calculate such a contribution to arbitrary orders, one needs control over the colour structure of arbitrarily complicated final states.
  
  As previously mentioned, one of the on-going areas of investigation in order to understand the origin of the super-leading logarithm terms is a study of the choice of ordering variable. Although we present here results obtained using transverse momentum, our approach only requires this to be specified at a late stage of the calculation. The full set of ordered diagrams could be used with a series of integrals calculated in a different ordering scheme in order to test the assumptions made.
  
  The presence of super-leading terms in other non-global observables is expected and, as for the ordering variable, the ordered diagram sets remain valid.  To use these one must relate the structure of the diagrams to the integrals for the observable required.
  
  \section*{Acknowledgements}
  We thank Jeff Forshaw and Simone Marzani for interesting discussions and help cross-checking the results. JK gratefully acknowledges financial support from the UK STFC.

  \appendix
  \section{C++ Implementation}

\subsection{Forming Uncut Diagrams}
The unique set of uncut diagrams was achieved by iteratively adding a gluon in all possible ways to the previous order set of diagrams. To ensure only topologically distinct sets were retained we used the $C++$ standard template library $set$ template. For this we needed to define a unique numbering system for each diagram, and a less-than operator, which allowed the $set$ functions to identify unique diagrams.

The gluons were numbered as follows. Gluons attached to one of the hard partons were numbered first, starting at one, increasing from left to right along the hard lines in turn. Then internal gluons are dealt with, the lowest numbers remaining are assigned to the gluons joining a vertex containing the lowest already numbered gluon, e.g. if $n$ gluons attached to hard partons, two gluons meeting gluon 1 at a triple gluon vertex would be assigned the numbers $n+1$ and $n+2$. To decide which of the two gluons receives which number, the lowest number is assigned to the gluon closest to gluon 1 (i.e. with fewest vertices in between). If the two gluons are equidistant from gluon 1 (as in our previous example) their distances to gluon 2 are compared and so on until they can be separated. This continues until all gluons are numbered.

 \FIGURE[ht]{
\label{numbering}
\centering
\SetWidth{2}
\SetScale{1}
\begin{centering}
\resizebox{3.5in}{!}{
\begin{picture}(240,150)(0,100)
\SetColor{Black}
%
\ArrowLine(0,250)(240,250)
\ArrowLine(0,100)(240,100)
\SetColor{Blue}
\Gluon(120,250)(120,220){4}{2}
\Gluon(40,100)(60,130){4}{2}
\Gluon(200,100)(180,130){-4}{2}
\Gluon(60,130)(80,160){4}{2}
\Gluon(80,160)(100,190){4}{2}
\Gluon(100,190)(120,220){4}{2}
\Gluon(180,130)(160,160){-4}{2}
\Gluon(160,160)(140,190){-4}{2}
\Gluon(140,190)(120,220){-4}{2}
\Gluon(100,190)(140,190){4}{3}
\Gluon(160,160)(80,160){-4}{6}
\Gluon(60,130)(180,130){4}{7}

\SetColor{Black}
\put(110,235){1}
\put(37,115){2}
\put(200,115){3}
\put(97,205){4}
\put(140,205){5}
\put(57,145){6}
\put(115,117){7}
\put(180,145){8}
\put(77,175){9}
\put(115,177){10}
\put(160,175){11}
\put(115,147){12}
\SetColor{Green}
%
%
%
\end{picture}
}
\end{centering}
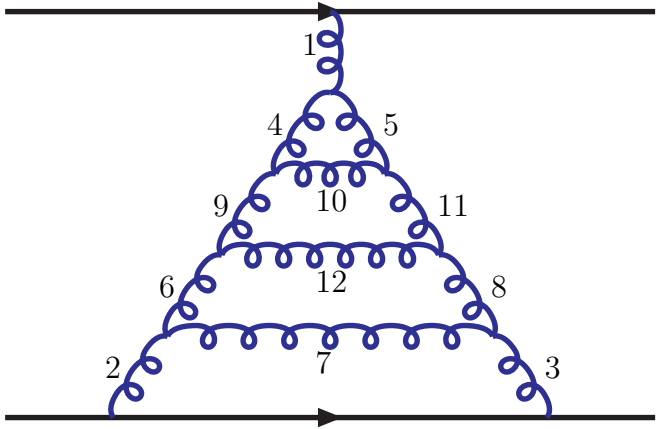
\caption{A fifth order diagram numbered according to our scheme.}
}

We now have a unique descriptor of each diagram consisting of a list of lines with numbered gluon connections, and a list of triple gluon vertices, each with three attached gluons. An example is shown in figure~\ref{numbering}. First gluons 1, 2 and 3 are numbered, as they join hard lines. Then the vertex containing gluon 1 has the next lowest numbers attached, 4 on the line closest to 2 (three vertices separate the two), 5 on the remaining line. After that the vertex joining 2 is done, with 6 numbering the line closest to 1. Next we treat the vertex containing 3, gluon 7 is already assigned so the remaining gluon is labelled 8. We then apply the same algorithm to with the vertices containing 4, 5, 6 and all gluons are numbered. The diagram in figure~\ref{numbering} would be stored in the program as:

Line 0: 1

Line 1: 2 3

Vertex 0: 1 4 5

Vertex 1: 2 6 7

Vertex 2: 3 7 8

Vertex 3: 4 9 10

Vertex 4: 5 10 11

Vertex 5: 6 9 12

Vertex 6: 8 11 12

The number of diagrams we would expect, corresponding to our constraints, could not be trivially calculated. However we compared numbers for subsets of diagrams (e.g. two and three parton states) using the \textsf{qgraf} program\cite{Nogueira:1991ex} and found agreement in all the areas we checked.

At this stage in the code, the colour factor of each uncut diagram was calculated via an interface to the \textsf{COLOUR} program \cite{Hakkinen:1996bb}. For the triple gluon vertices we simply entered the connections in increasing numerical order, as stored by the code's internal numbering system. The factors of $-1$ originating in the anti-symmetric nature of the triple gluon coupling were resolved later, as described in section~\ref{minus1b}. The colour factors are stored as polynomials in $N$
  
  \subsection{Cutting and Ordering Diagrams}
  The full set of cut and ordered diagrams was obtained by systematically producing all cuts and orders and discounting those that failed our requirements. Connections on hard lines and triple gluon vertices were simply assigned in turn to either the left or right of the cut. Diagrams with disconnected gluons were thrown away at this stage.
  
  The ordering constraints were kept track of through a series of theta functions, each describing a less than-greater than relation. Moving outwards from the hard process along the hard parton lines strong ordering was recorded, when diagrams produced contradictions they were thrown away. Then for remaining diagrams the triple gluon vertices were treated in turn, with one gluon chosen as the softer gluon, and the remaining two joining to form the hard emitted gluon. Resulting constraints were added at each stage and diagrams discarded as soon as any inconsistencies were found.
  
  When all triple gluon vertices had been dealt with, the result of the fully ordered diagram was stored as a list of gluons with information on which hard partons or other gluons they connected, and whether they were virtual or real.

  \subsection{Integrals}\label{appint}
  In the high energy limit, there are very few required integrals, and they were calculated by hand (see section~\ref{intsect}) and only the colour factors and rapidity integrals change between diagrams. The definition of the observable we calculate is most easily described in terms of a cone algorithm with radius $R$, but in fact since we use an approximation in which all emissions are assumed to be well separated in rapidity our results are general for any jet algorithm.

 The main complication that exists is that the rapidity integral limits when there are many triple gluon connections lead to a series of complex nested integrals, which are dealt with as follows.
  
  A series of theta (step) functions describing the rapidity limits are taken, describing which of the rapidity regions illustrated in figure~\ref{etaphi} the gluons fall into. 
  
  So, for example, from equation~(\ref{sigma2}) we would retain
\begin{equation}\label{theta1}
C(N)\left|\theta\left(\frac{\delta\eta}{2} -y_1\right)\theta\left(y_1+\frac{\delta\eta}{2}\right)\right|_{y_1=virt}\left|\theta\left(\frac{\delta\eta}{2} -y_2\right)\theta\left(y_2+\frac{\delta\eta}{2}\right)\right|_{y_2=real}
\end{equation}
plus an imaginary part. Terms not noted here are common to all diagrams at this order.

The virtual gluon produces a term $Y+2R$ (where $R=\frac{\delta\eta-Y}{2}$ is an effective jet radius) and the real gluon a term $2R$ (being vetoed in region $Y$). So the total rapidity dependent part is $2RY+4R^2$. 
  
  If a gluon lies between, for example, $-\delta\eta/2$ and $\bar{Y}/2$, it is split into two terms, one with an in-gap gluon and one with an out-of-gap gluon
  \begin{equation}\label{thetaexpand}
  \theta\left(\frac{\bar{Y}}{2} -y_1\right)\theta\left(y_1+\frac{\delta\eta}{2}\right)\to\theta\left(\frac{\bar{Y}}{2} -y_1\right)\theta\left(y_1-\frac{\delta\eta}{2}\right)+\theta\left(\frac{\delta\eta}{2} -y_1\right)\theta\left(y_1+\frac{\delta\eta}{2}\right),
  \end{equation}
  and the resulting terms are $Y+2R$ and $\bar{Y}$.
  
  Gluons may also be re-emitted, and therefore limited by the rapidity of the emitting gluon, which produces nested integrals. For example 
  \begin{equation}\label{thetanest}
  \theta\left(\frac{Y}{2}-y_1\right) \theta\left(\frac{Y}{2}+y_1\right)  \theta\left(\frac{Y}{2}-y_2\right) \theta\left(\frac{Y}{2}+y_2\right)\theta(y_1-y_2)
\end{equation}
  gives $Y^2/2!$. To deal with all cases of this kind, we expand the theta functions, so for example
   \begin{eqnarray}\label{thetanest2}
  \nonumber\theta\left(\frac{Y}{2}-y_1\right) \theta\left(\frac{Y}{2}+y_1\right)  \theta\left(\frac{Y}{2}-y_2\right) \theta\left(\frac{Y}{2}+y_2\right) \\
 \nonumber\to\theta\left(\frac{Y}{2}-y_1\right) \theta\left(\frac{Y}{2}+y_1\right)  \theta\left(\frac{Y}{2}-y_2\right) \theta\left(\frac{Y}{2}+y_2\right)\theta(y_1-y_2)\\
 +\theta\left(\frac{Y}{2}-y_1\right) \theta\left(\frac{Y}{2}+y_1\right)  \theta\left(\frac{Y}{2}-y_2\right) \theta\left(\frac{Y}{2}+y_2\right)\theta(y_2-y_1)
\end{eqnarray}
giving a result of $Y^2/2+Y^2/2=Y^2$ as we would expect for two gluons spanning the gap region. However this expansion has the advantage of automatically including all term of the form of eq.~(\ref{thetanest}).

With these fully expanded set of theta functions the rapidity dependent integral is the width of a rapidity region raised to the power of the number of gluons in that region, over the factorial of that number of gluons,  summed over all rapidity regions:
\begin{equation}\label{rapint}
\sum_{i} \frac{(\Delta y)_i^{n_i}}{{n_i}!}
\end{equation}
where $(\Delta y)_i$ are the width of the regions, $i$, (e.g. $Y$, $R$, $\bar{Y}$) and $n_i$ is the number of gluons lying within each region.

This simple formula holds for all diagrams after the theta function expansion has taken place. The performance suffers due to the large increase in the number of terms to be dealt with, but the benefits are the simplicity of the integration, and being sure we have systematically included all the required terms.

\subsection{Performance and Parallelisation}
The performance of the code decreases rapidly as the $O(\alpha_s)$, and therefore the number of diagrams, increases. The table below shows the number of diagrams at three stages of the program, and the approximate running time (including compilation time).

\vspace{2ex}

\begin{tabular}{|c|l|l|l|l|}
\hline
$O(\alpha_s)$ & Uncut diagrams & Cut diagrams & Ordered diagrams & Run time(s)\\
\hline
1 & 6 & 24 & 24 & ~21 \\
\hline
2 & 55 & 564 & 672 & ~22\\
\hline
3 & 681 & 17,194 & 30,048 & ~44\\
\hline
4 & 10,529 & 529,910 & 1,746,272  & ~3700\\
\hline
5 & 193,741 & 22,521,544 & 127,424,961 & ~1,600,000\\
\hline
\end{tabular}

\vspace{2ex}

It can be seen that the processing time increases at a higher rate than the number of uncut diagrams, as the number of cut diagrams, possible orderings of those diagrams and (not shown) expansions to account for nested integrals (section~\ref{appint}) all increase per-diagram with order.

Although times were manageable up to fourth order, to reach fifth order it was necessary to run the program in parallel sessions. As the uncut diagrams need to be checked to ensure they form a unique set, they must be created for each order as one stage. The uncut connections were then written to a series of text files. The connections were constant for any $2\to2$ process, so this stage only need be done once. 

Each file was read in and the pre-prepared connections combined with a 0th order diagram ($qq\to qq$, $gg\to gg$ etc) and colour factors calculated. From here we have the normal set of uncut diagrams and the code can proceed as usual with the following steps. Results in the form of a colour factor (polynomial in $N$) and term in $Y$, $\bar{Y}$, $i\pi$ and $R$ are then written to a text file. The program can then read in these results files, add them and output the result.

A similar process can be used within each run of the program. The uncut diagram set is split into smaller sets,  and these sets run through the intermediate stages to a result. A small number of uncut diagrams can quickly give rise to a large number of cut diagrams, ordered diagrams and integral terms, which can cause problems in temporary memory, before collapsing down to a small number of result terms. Subdividing allows the full result for the larger set of uncut diagrams to be calculated without at any one time storing a large number of the diagrams necessary for the intermediate stages.

\bibliographystyle{JHEP}
\bibliography{Mybibliography}

\providecommand{\href}[2]{#2}\begingroup\raggedright\begin{thebibliography}{10}

\bibitem{Kidonakis:1998nf}
N.~Kidonakis, G.~Oderda, and G.~Sterman, {\it {Evolution of color exchange in
  {QCD} hard scattering}},  {\em Nucl. Phys.} {\bf B531} (1998) 365--402,
  [\href{http://xxx.lanl.gov/abs/hep-ph/9803241}{{\tt hep-ph/9803241}}].

\bibitem{Oderda:1998en}
G.~Oderda and G.~Sterman, {\it {Energy and color flow in dijet rapidity gaps}},
   {\em Phys. Rev. Lett.} {\bf 81} (1998) 3591--3594,
  [\href{http://xxx.lanl.gov/abs/hep-ph/9806530}{{\tt hep-ph/9806530}}].

\bibitem{Oderda:1999kr}
G.~Oderda, {\it {Dijet rapidity gaps in photoproduction from perturbative
  {QCD}}},  {\em Phys. Rev.} {\bf D61} (2000) 014004,
  [\href{http://xxx.lanl.gov/abs/hep-ph/9903240}{{\tt hep-ph/9903240}}].

\bibitem{Berger:2001ns}
C.~F. Berger, T.~Kucs, and G.~Sterman, {\it {Energy flow in interjet
  radiation}},  {\em Phys. Rev.} {\bf D65} (2002) 094031,
  [\href{http://xxx.lanl.gov/abs/hep-ph/0110004}{{\tt hep-ph/0110004}}].

\bibitem{Appleby:2003sj}
R.~B. Appleby and M.~H. Seymour, {\it {The resummation of inter-jet energy flow
  for gaps-between- jets processes at HERA}},  {\em JHEP} {\bf 09} (2003) 056,
  [\href{http://xxx.lanl.gov/abs/hep-ph/0308086}{{\tt hep-ph/0308086}}].

\bibitem{Dasgupta:2001sh}
M.~Dasgupta and G.~P. Salam, {\it {Resummation of non-global QCD observables}},
   {\em Phys. Lett.} {\bf B512} (2001) 323--330,
  [\href{http://xxx.lanl.gov/abs/hep-ph/0104277}{{\tt hep-ph/0104277}}].

\bibitem{Dasgupta:2002bw}
M.~Dasgupta and G.~P. Salam, {\it {Accounting for coherence in interjet E(t)
  flow: A case study}},  {\em JHEP} {\bf 03} (2002) 017,
  [\href{http://xxx.lanl.gov/abs/hep-ph/0203009}{{\tt hep-ph/0203009}}].

\bibitem{Forshaw:2006fk}
J.~R. Forshaw, A.~Kyrieleis, and M.~H. Seymour, {\it {Super-leading logarithms
  in non-global observables in QCD}},  {\em JHEP} {\bf 08} (2006) 059,
  [\href{http://xxx.lanl.gov/abs/hep-ph/0604094}{{\tt hep-ph/0604094}}].

\bibitem{Forshaw:2008cq}
J.~R. Forshaw, A.~Kyrieleis, and M.~H. Seymour, {\it {Super-leading logarithms
  in non-global observables in QCD: Colour basis independent calculation}},
  {\em JHEP} {\bf 09} (2008) 128,
  [\href{http://xxx.lanl.gov/abs/0808.1269}{{\tt arXiv:0808.1269}}].

\bibitem{Catani:1985xt}
S.~Catani, M.~Ciafaloni, and G.~Marchesini, {\it {NONCANCELLING INFRARED
  DIVERGENCES IN QCD COHERENT STATE}},  {\em Nucl. Phys.} {\bf B264} (1986)
  588--620.

\bibitem{Bonciani:2003nt}
R.~Bonciani, S.~Catani, M.~L. Mangano, and P.~Nason, {\it {Sudakov resummation
  of multiparton QCD cross sections}},  {\em Phys. Lett.} {\bf B575} (2003)
  268--278, [\href{http://xxx.lanl.gov/abs/hep-ph/0307035}{{\tt
  hep-ph/0307035}}].

\bibitem{Forshaw:2005sx}
J.~R. Forshaw, A.~Kyrieleis, and M.~H. Seymour, {\it {Gaps between jets in the
  high energy limit}},  {\em JHEP} {\bf 06} (2005) 034,
  [\href{http://xxx.lanl.gov/abs/hep-ph/0502086}{{\tt hep-ph/0502086}}].

\bibitem{Forshaw:2007vb}
J.~R. Forshaw and M.~Sjodahl, {\it {Soft gluons in Higgs plus two jet
  production}},  {\em JHEP} {\bf 09} (2007) 119,
  [\href{http://xxx.lanl.gov/abs/0705.1504}{{\tt arXiv:0705.1504}}].

\bibitem{Nogueira:1991ex}
P.~Nogueira, {\it {Automatic Feynman graph generation}},  {\em J. Comput.
  Phys.} {\bf 105} (1993) 279--289.

\bibitem{Hakkinen:1996bb}
J.~Hakkinen and H.~Kharraziha, {\it {Colour: A computer program for QCD colour
  factor calculations}},  {\em Comput. Phys. Commun.} {\bf 100} (1997)
  311--321, [\href{http://xxx.lanl.gov/abs/hep-ph/9603229}{{\tt
  hep-ph/9603229}}].

\end{thebibliography}\endgroup

  \end{document}